\documentclass[aps, showpacs, pra, superscriptadaddress, 
twocolumn] {revtex4-1}
\usepackage{hyperref}
\usepackage{amsmath }
\usepackage{amssymb}
\usepackage{epsfig}
\usepackage{natbib}
\usepackage{epstopdf}
\usepackage{graphicx}

\usepackage{floatrow}
\usepackage[caption=false,font=Large]{subfig}

\newcommand*{\be}{\begin{equation}}
\newcommand*{\ee}{\end{equation}}
\newcommand*{\bea}{\begin{eqnarray}}
\newcommand*{\eea}{\end{eqnarray}}

 \DeclareFontFamily{OT1}{pzc}{}
 \DeclareFontShape{OT1}{pzc}{m}{it}%
 {<->  s  *  [1.400]  pzcmi7t}{}
\DeclareMathAlphabet{\mathscr}{OT1}{pzc}%
{m}{it}

\begin{document}

\title{Blow-up regimes in  the $\mathcal{PT}$-  and the $\mathcal{AC}$-dimer }

\author{I  V Barashenkov$^{1,2,3}$,  G  S Jackson$^2$, and S Flach$^1$}
 \affiliation{
$^1$  New Zealand Institute for Advanced Study, 
 Centre for Theoretical Chemistry and Physics, Massey University, Auckland 0745, New Zealand 
 \\  $^2$  
 Department of Mathematics and Centre for Theoretical  and Mathematical Physics,  University of Cape Town, Rondebosch 7701, South Africa 
  \\  $^3$  
 Joint Institute for Nuclear Research, Dubna, Russia}

\begin{abstract}
In the actively coupled ($\mathcal{AC}$)  pair of waveguides,
 the growth of small perturbations is saturated by the focussing nonlinearity that  couples the linearly growing to the 
linearly damped mode. On the other hand, in the $\mathcal{PT}$-symmetric coupler,
 the focussing nonlinearity  promotes 
the blowup of stationary light beams.
The purpose of this study is to compare the nonlinear dynamics 
  and explain the opposite effect of the same nonlinearity in the two systems.
We show that while the blowup regimes
are stable in the 
$\mathcal{PT}$-symmetric pair of waveguides, 
they are unstable and hence 
cannot be observed in the $\mathcal{AC}$-dimer.
\end{abstract}

\pacs{}
\maketitle

\section{Introduction} 

The current growth of interest in the $\mathcal{PT}$-symmetric photonic systems with gain and loss
\cite{PT_breaking,SXK,RKEC,Musslimani,Zheng,Longhi,Guo,Ramezani,Rueter,Recent_PT,BSSDK} is motivated by the 
unusual  phenomenology associated with these systems.
Optical structures composed of coupled active and lossy elements 
exhibit 
symmetry-breaking phase transitions \cite{SXK,Rueter,PT_breaking},
 unconventional beam refraction \cite{Musslimani,Zheng}, 
nonreciprocity \cite{Longhi,Rueter}, 
  loss-induced  transparency \cite{Guo},
conical diffraction \cite{Ramezani},
and beam breathing \cite{SXK,Musslimani,Rueter,BSSDK}.
 The nonlinear effects in such systems 
can be utilised  for an efficient control of light, including
all-optical low-threshold switching \cite{RKEC,SXK}
and unidirectional
invisibility \cite{SXK,RKEC}.
One of the two objects considered in
the present paper is the simplest $\mathcal{PT}$-symmetric optical system
consisting of a single waveguide with loss coupled to a waveguide with an 
equal amount of gain. 

The gain-loss systems ---
and in particular the $\mathcal{PT}$-symmetric coupler we discuss here --- 
display a variety of dynamical regimes, including stationary,
periodic, as well as blow-up regimes where 
the power in one of the waveguides grows without bound.
The blow-up is obviously an undesirable effect
in an optical system.
 In this paper, we
study the blowing up regimes  of the $\mathcal{PT}$-symmetric
coupler,
and compare  them to  dynamical regimes in another 
finite-dimensional system with gain and loss:
the actively coupled ($\mathcal{AC}$) pair of wavegides.

The $\mathcal{AC}$-dimer was proposed as a configuration of gain and loss
alternative to the $\mathcal{PT}$-symmetric coupler.
Mathematically, the system  can be  shown to have
 a blow-up solution; however this regime is 
not observed in the numerical simulations of the system \cite{ABRF}.
Instead,  generic initial conditions  set off  an exponential growth of a linearly excitable mode                             
which is then  saturated by the nonlinear coupling of this mode to an energy-draining mode.
As a result, all dynamical regimes observed in the $\mathcal{AC}$ coupler are bounded
\cite{ABRF}. 

The issue that concerns us here, is why  this mechanism is
not at work in the case of the $\mathcal{PT}$-dimer ---  that is, why
 does the same, focussing Kerr, nonlinearity not couple the 
growing to the damped mode there.

We show that the answer is in the geometry of the corresponding phase spaces.
 The phase space of the $\mathcal{PT}$-dimer is foliated into
coaxial cylinders.
Despite the presence of gain and loss, the motion on 
each (two-dimensional) cylindric surface
 is conservative,
 with the gain-loss terms producing an inverted harmonic oscillator  potential
 which sends the power to infinity.
   The nonlinearity gives rise to finite-depth wells in the potential, but cannot
  eliminate the negative potential as a whole.
  The potential wells harbour periodic motions of the dimer;
  however the blow-up regimes remain available for any value of the gain-loss coefficient. 
  There are continuous families of blowing-up trajectories, lying on cylinders of different radius.
A small perturbation may push the phase point from one cylinder to  another, but
this will simply amount to the transition from one family of unbounded trajectories to 
another.
 
On the other hand, the phase space of the $\mathcal{AC}$ dimer is three-dimensional. 
There are continuously many  blowing-up trajectories, but they are all asymptotic to the vertical axis.
Because this funnel of raising trajectories becomes exponentially thin as 
$Z \to \infty$, the blow-up is unstable. For a sufficiently large $Z$, a small perturbation
in the horizontal plane ``knocks" the trajectory out of the funnel. The 
trajectory is then captured into a limit cycle or a strange attractor, i.e. remains in the 
finite part of the space.

The outline of this paper is as follows.
The ${\mathcal PT}$ coupler is considered in 
 section \ref{sPT}.
After producing a
particular explicit blow-up solution, we elucidate the cylindrical foliation of the phase space, 
provide an effective-particle description of trajectories on the cylindrical surfaces, 
 and
 classify fixed points.  In the symmetry-broken phase, the system-dynamic analysis is supplemented with 
 the demonstration of the blow-up on the basis of the power-imbalance estimates.
In section \ref{sAC}, we turn to the ${\mathcal AC}$ dimer.
 We first prove that the defocusing nonlinearity cannot 
 arrest the growth of linear perturbations
 and hence presents no alternative to the ${\mathcal PT}$-symmetric model. 
After that we analyse the phase space of the $\mathcal{AC}$-coupler with the 
focussing nonlinearity and prove instability of its blowup regime. 
Section \ref{Conclusions} summarises our results for the two types of dimers
and draws conclusions.

\section{$\mathcal{PT}$-symmetric dimer}
\label{sPT}

The  nonlinear coupler with gain and loss
was proposed in \cite{CSP}, as an improvement of the conventional 
twin core coupler. More recently
this optical configuration  has attracted attention
as an experimentally  realisable  $\mathcal{PT}$-symmetric
system \cite{Guo,Rueter,RKEC,SXK}.

The structure consists of two optical waveguides in close proximity 
to one another. One guide has a certain amount of loss and the other 
one an equal amount of optical gain. 
The corresponding mode amplitudes satisfy 
\begin{subequations}
\label{coupler}
\begin{align}
i \frac{d \psi_1}{dz}   + |\psi_1|^2 \psi_1 +\psi_2  &  =  i \gamma \psi_1, \\
i \frac{d \psi_2}{dz}  + |\psi_2|^2 \psi_2 + \psi_1 &  =   -i \gamma \psi_2.
\end{align}
\end{subequations}
Here $z$ stands for the 
distance along the guide while $\gamma>0$ is the gain-loss coefficient.
The quantities $P_1=|\psi_1|^2$ and $P_2=|\psi_2|^2$ measure  the power carried by the
active and the lossy mode, respectively.

The two-wire $\mathcal{PT}$-symmetric coupler can be seen as the simplest  finite 
chain of symmetrically balanced waveguides with gain and loss \cite{finite}, 
or the elementary constituent of an infinite chain \cite{infinite}.

Note that the sign of the nonlinearity can be chosen arbitrarily in 
the equations of the $\mathcal{PT}$-symmetric dimer.
Indeed, the system with the opposite sign of the nonlinear term,
\begin{subequations}
\label{defoc}
\begin{align}
i \frac{d \varphi_1}{dz}   - |\varphi_1|^2 \varphi_1 +\varphi_2  &  =  i \gamma \varphi_1, \\
i \frac{d \varphi_2}{dz}  - |\varphi_2|^2 \varphi_2 + \varphi_1 &  =   -i \gamma \varphi_2.
\end{align}
\end{subequations}
can be mapped to \eqref{coupler}  by
 the ``staggering" transformation
\be
\psi_1= - \varphi_1^*, \quad \psi_2= \varphi_2^*.
\label{stag}
\ee
Therefore,  the focussing and defocussing nonlinearity
are equivalent and
we can restrict ourselves to considering the dimer in the form \eqref{coupler}.

The $\mathcal{PT}$ symmetry manifests itself as 
the invariance with respect to the permutation $\psi_1 \rightleftarrows \psi_2$ followed by
taking 
the complex conjugates of $\psi_1$, $\psi_2$, and the ``time" inversion: $z \to -z$. 
When $\gamma > 1$, small-amplitude inputs grow exponentially; it is customary 
to say that the $\mathcal{PT}$-symmetry is spontaneously broken. 
On the contrary, when $\gamma \leq 1$, the $\psi_{1,2}=0$ solution is stable; 
the  symmetry is said to be exact, or unbroken.

The foundations of the mathematical analysis of  Eq.\eqref{coupler} 
were laid 
in \cite{RKEC}
where 
the $\mathcal{PT}$-symmetric dimer was shown to 
define a completely integrable system. 
However no explicit solutions were found so far, and the dynamics had to be 
analysed numerically \cite{RKEC,SXK}.
The numerical simulations have revealed the coexistence of the blow-up regimes, where 
the total power $|\psi_1|^2+|\psi_2|^2$ grows without bound,
with periodic trajectories
 \cite{RKEC,SXK}.
 
 In a very recent communication \cite{KPT}, its authors 
 have established several additional properties of solutions to \eqref{coupler}.
 In particular, they proved (i) that  solutions do not blow up in finite time; (ii)
 that in the symmetry-unbroken phase ($\gamma<1$) 
 small-amplitude solutions remain bounded for all times
 but (iii) there are large-amplitude solutions that 
grow exponentially fast. Our approach is different from the one 
in \cite{KPT} and our results in this section complement those in \cite{KPT}.

\subsection{Explicit blowup solution}

A particular blow-up solution can be found explicitly
--- both for $\gamma>1$ and $\gamma <1$. 
Introducing $p$ and $q$ by
\[
\psi_1(z)= e^{\gamma z} p(z), \quad \psi_2(z)= e^{-\gamma z} q(z),
\]
and defining $\eta= e^{\gamma z}$,  Eqs.\eqref{coupler} become
\begin{subequations}
\label{pq}
\begin{align}
i \gamma p_\eta + \eta|p|^2p + \eta^{-3}q=0, \\
i\gamma q_\eta  + \eta^{-3} |q|^2q+ \eta p=0.
\end{align}
\end{subequations}
Assuming now that the complex fields $p$ and $q$ have a common phases: $p=a e^{i \phi}$ and $q=b e^{i \phi}$,
and substituting in \eqref{pq},  we conclude that $a$ and $b$ are constant, with $ab=1$,
and that 
\[
\phi= \frac{1}{2\gamma} (a^2 \eta^2- \frac{1}{a^2 \eta^2}).
\]
This gives an 
exact blow-up solution to the $\mathcal{PT}$-symmetric coupler:
\begin{subequations}
\label{A4}
\begin{align}
\psi_1(z)= 
\exp \{
\gamma(z-z_0)+ \frac{i}{\gamma}  \sinh [ 2 \gamma(z-z_0)]\},
\label{A5} \\
\psi_2(z)= \exp \{ -\gamma(z-z_0) + \frac{i}{\gamma} \sinh [ 2\gamma(z-z_0)]\},
\label{A6}
\end{align}
\end{subequations}
where we have defined $z_0$ such that $a=e^{-\gamma z_0}$. 
The constant $z_0$ is a free parameter in \eqref{A4}
which results from the translation invariance of Eqs.\eqref{coupler}.

The existence of an unbounded
trajectory in the $\gamma \leq 1$ region
does not contradict the 
stability of  the $\psi_{1,2}=0$ 
solution here.
Indeed,  the solution \eqref{A4} does not have a small-amplitude limit:
it tends to zero neither as $z \to -\infty$ nor as $z \to \infty$.

\subsection {Cylindrical phase space foliation}
\label{traj} 

To obtain the general solution of equations \eqref{coupler}
and understand the geometry of the phase space, we reformulate these
   \cite{CSP,RKEC}
 in terms of the Stokes variables
\begin{align} 
 X=\frac12 (\psi_1 \psi_2^*+ \psi_1^*\psi_2), \nonumber  \\
 Y=\frac{i}{2} (\psi_1 \psi_2^*-\psi_1^*\psi_2),    \nonumber \\
  Z=\frac12 (|\psi_1|^2-|\psi_2|^2).
 \label{change}
 \end{align}
  Eqs.\eqref{coupler} then acquire the form
 \begin{subequations} \label{Z1}
 \begin{align}
 {\dot X}=YZ, \\
 {\dot Y}= Z(1-X), \\
 {\dot Z}= \gamma r-Y,   \label{J1}
 \end{align}
 \end{subequations}
 where $r=\sqrt{X^2+Y^2+Z^2}$
 and the dot stands for the derivative with respect to $t=2z$.
 
 Note that despite the equations \eqref{coupler} governing four independent 
 real variables, the system \eqref{Z1} is only for three unknowns.
 The equation for the phase of $\psi_1$  
 decouples from the rest of the dynamical system \eqref{Z1} which
 involves  the difference of the phases of $\psi_1$ and $\psi_2$
  but not the phases themselves.
Letting $\psi_1=\sqrt{P_1}e^{i \Phi_1}$, 
 we have
 \[
 {\dot \Phi_1}= \frac12 (r+Z)+\frac{X}{2(r+Z)}.
 \]
 Therefore the dynamics described by the system \eqref{coupler} 
 are effectively three-dimensional. We now show that in fact, 
 all its trajectories lie on two-dimensional surfaces.

\begin{figure}
  \sidesubfloat[]{
  \includegraphics[width=0.37\linewidth]{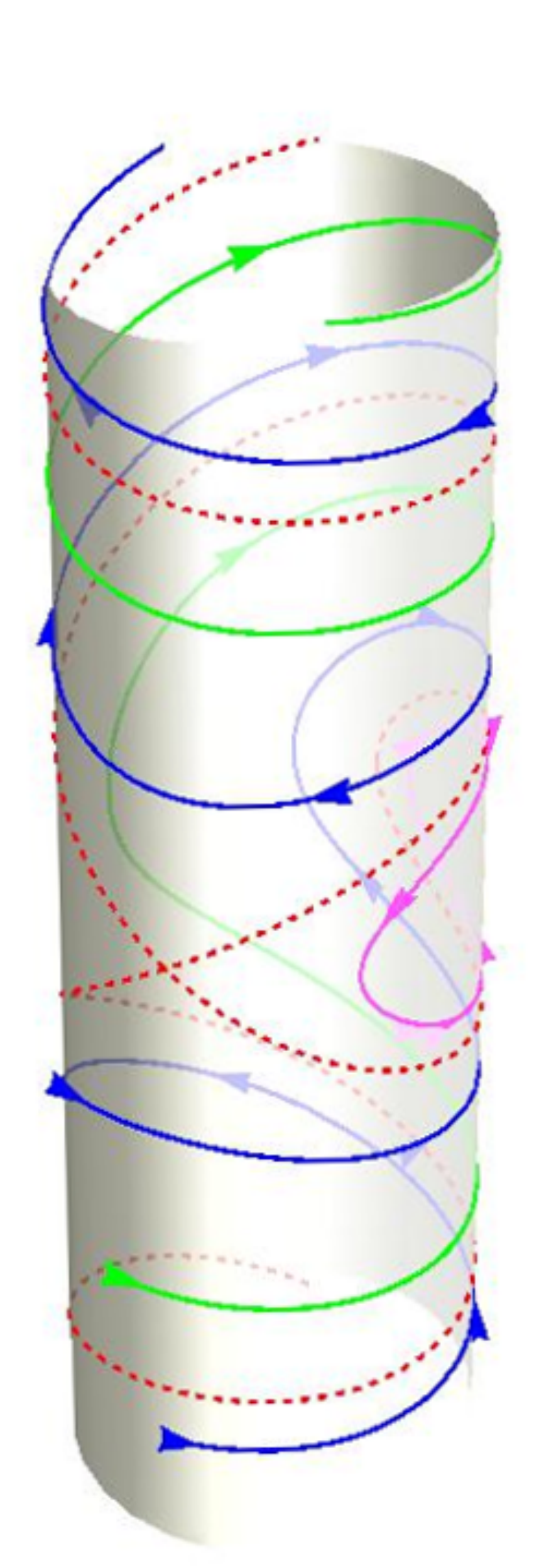}\label{fig:sub1}
  }
  \sidesubfloat[]{
  \includegraphics[width=0.46\linewidth]{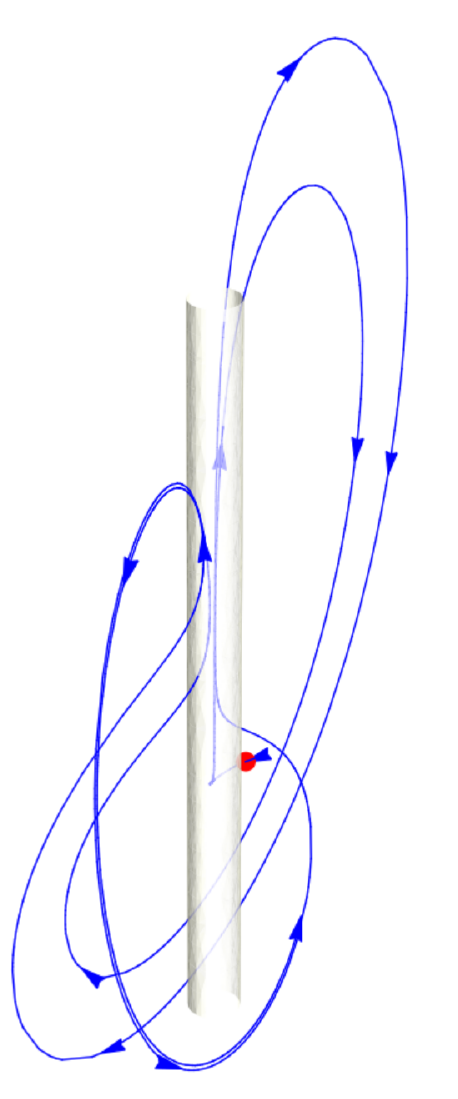}\label{fig:sub2}
  } 
  \caption{(Color online) --- 
    (a) Representative trajectories
    of the $\mathcal{PT}$-symmetric coupler
    on the surface of a cylinder.
      A periodic trajectory ---
     the closed solid curve in the middle of the cylinder, plotted in purple (gray)  --- is confined
    within the separatrix loop (plotted in dashed red).
 The other two solid curves --- dark blue (black) and green (light gray)  ---
  wind up to $Z \to \infty$ 
    and represent blow-up regimes of the coupler.
     Here $\gamma=0.5$ and $\rho =3$.
    (b) A representative trajectory of the $\mathcal{AC}$-coupler.
    From the point shown by the red blob,
    the trajectory zaps onto the vertical axis, starts moving up, 
    but gradually deviates from the  vertical and 
    leaves the imaginary  cylinder of small radius.
    The trajectory ends up approaching a limit cycle. The blowup is arrested.
    Here $\gamma = 1.9$, $a=2$ and the initial
    conditions $(X,Y,Z) = (0,5,0)$.
    \label{cylinder}
  }
\end{figure}

 Transforming to  the 
 cylindrical polars
 \[
 X=1+ \rho \sin \theta, \quad
 Y = \rho \cos \theta,
 \]
 where $\rho \geq 0$ and $0 \leq \theta < 2 \pi$, 
  Eqs.\eqref{Z1} yield
 \be
 {\dot \theta}  = Z \label{Z5}   
\ee
and
\be
{\dot Z}=\gamma r - \rho \cos \theta,  \label{Z4}
\ee
where
\be
r= \sqrt{Z^2+2 \rho \sin \theta + \rho^2+1}.
\label{Z40}
\end{equation}
The third  equation is ${\dot \rho=0}$ which
 implies that $\rho$ is constant:  the motion is always on a
cylindrical surface
(see Fig.\ref{cylinder}(a)).  

Differentiating \eqref{Z40} and using 
 \eqref{Z4} and \eqref{Z5} we obtain
\be
{\dot r}= \gamma Z.
 \label{Z7}
\ee
 Comparing this to \eqref{Z5}, we get
 ${\dot r}= \gamma {\dot \theta}$, whence 
 \be
 r= \gamma (\theta -\chi).
 \label{Z8}
 \ee
 Here the constant $\chi$ is defined by the initial conditions $\rho(0), Z(0)$
 and $\theta(0)$:
 \be
 \chi=\theta(0)-\frac{1}{\gamma} \sqrt{ Z^2(0)+2 \rho(0) \sin \theta(0) + \rho^2(0)+1}. 
 \label{Y1}
 \ee


 \floatsetup[figure]{style=plain,subcapbesideposition=top}

 Using \eqref{Z8},  Eqs.\eqref{Z5} and \eqref{Z4} become
  \begin{align}
{\dot \theta}  = Z,  \quad
{\dot Z}=\gamma^2 (\theta -\chi) - \rho \cos \theta. 
\label{Z10}
\end{align} 
This has an obvious conservation law:
\be
{\dot \theta}^2 - \gamma^2 (\theta- \chi)^2 + 2 \rho \sin \theta +\rho^2+1=0,
\label{Z11}
\ee
where we used \eqref{Y1} to identify the constant of integration.
This is an equation for a curve on the surface of a cylinder of
 radius $\rho$. The curve is determined by the 
 angular parameter $\chi$. 

The cylindrical radius $\rho$ and the angle $\chi$ are two integrals of motion 
of the $\mathcal{PT}$-symmetric dimer \eqref{coupler}. The availability of two 
independent integrals makes the dimer a completely integrable system \cite{RKEC}.

The cylinder of the radius $\rho=0$ is exceptional. When $\rho=0$, $\theta$ is undefined
and Eq.\eqref{Z11} is invalid. However,  in this case
Eq.\eqref{J1} gives
${\dot Z}= \gamma \sqrt{Z^2+1}$, whence $Z= \sinh [ \gamma(t-t_0)]$.
This is  the trajectory corresponding to 
our explicit blowup solution \eqref{A4}.
(See also \cite{Pickton}.)

We note that equations similar to \eqref{Z10} were derived in \cite{KPT,Pickton}
within a different formalism.

\subsection{Imaginary particle representation}

Assume $\rho \neq 0$, and let  $\kappa= \gamma / \sqrt{2 \rho}>0$.
Denoting $\tau=\sqrt{\rho} t$ and 
 $q= \theta-\chi$,  Eq.\eqref{Z11}  acquires the form of 
 the energy conservation law for  a classical particle in the potential $V(q)$:
\be
 \frac{{q_\tau}^2}{2}     +V(q) = E.
\label{Z13}
\ee
Here 
\be
E=   -1-\frac{(\rho-1)^2}{2\rho}
\label{J7}
\ee
and
the potential 
\be
V(q)=-  \kappa^2 q^2  + \sin (q+\chi).
\label{V}
\ee
Since
 \[
\rho^2+ 2 \rho \sin(q+\chi)+1 \geq (\rho-1)^2 \geq 0,
\] 
it follows from 
Eq.\eqref{Z13} that  $|q_t| \leq \gamma |q|$. Letting $q=q(0) e^\phi$ yields $|\phi_t| \leq \gamma$,
and so
\be
|q(0)| e^{-\gamma t} \leq |q| \leq |q(0)| e^{\gamma t}.
\ee
This inequality implies that $q$ cannot grow faster than $e^{\gamma t}$.
(This result was previously established via the balance equations \cite{KPT}.)

According to \eqref{Y1}, adding a multiple of $2 \pi$ to $\theta$
changes $\chi$ but does not affect $q$.
Therefore without loss of generality 
 $\chi$ can be taken in the interval $(0, 2\pi)$.

 It is not difficult to realise that not all trajectories of the 
 imaginary particle correspond to evolutions of the 
 system \eqref{coupler}. 
 First of all,
 the value of $E$ in \eqref{J7} is bounded from above:
$E \leq -1$. Therefore only trajectories
of the particle  with $E \leq -1$ correspond to the 
dimer's trajectories on the surface of a cylinder with some $\rho$
(and $\gamma$ given by $\sqrt{2\rho} \kappa$).

 Second,
 in view of \eqref{Z8}, 
only positive $q$ represent configurations of the dimer.
 Any trajectory of the particle with $q$ reaching zero, or approaching zero as $\tau$ grows
 to infinity,  would correspond to a solution of the system \eqref{coupler} decaying to zero,
 in finite or infinite time: $|\psi_1|^2+|\psi_2|^2 \to 0$. 
 [In the next subsection, we will show that the 
 specific choice of the integration constant in \eqref{Z13} 
 is compatible with 
only one such trajectory.]

\begin{figure}
\begin{center}
  \includegraphics[width=\linewidth] {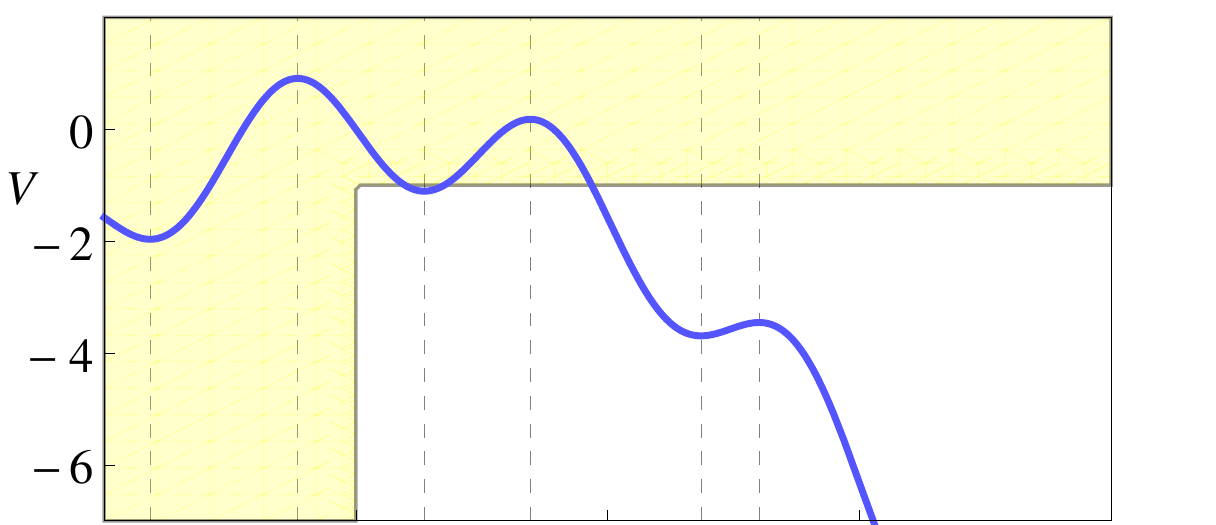}  
  \includegraphics[width=\linewidth] {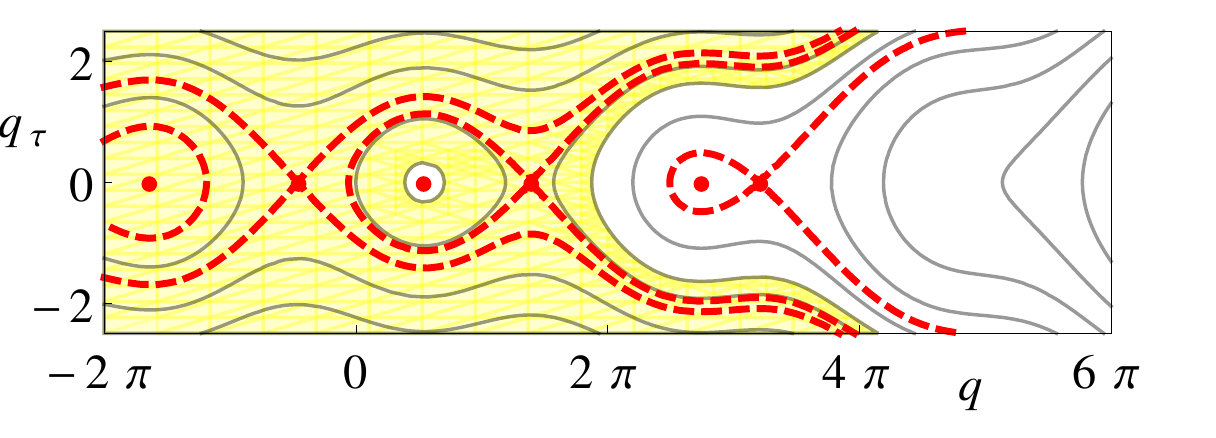}
\end{center}
 \caption{(Color online) --- 
    Top: the potential $V(q) = \sin (q + \chi) - \kappa^2 q^2$, 
    with the vertical lines indicating positions of the extrema.
     Here $\kappa = 0.2$ and $\chi = \pi$. 
     The negative $q$ and the potential values $V > -1$
     are inaccessible to the imaginary particle (shaded in yellow).
    Bottom:  the corresponding phase portrait with fixed points and separatrices
    shown in dashed  red, calculated via Eq.\eqref{Z13}.
    Shaded is  the region where $q<0$ or $E >-1$. Trajectories in this region do  not
    correspond to any motions of the dimer.
        \label{phase_portrait}
  }
\end{figure}

 The top panel in Fig.\ref{phase_portrait} sketches a potential $V(q)$
 for a particular set of values of $\kappa$ and  $\chi$. Shaded is the region where $q<0$ 
 and the section with $V >-1$.
  The bottom panel shows 
 trajectories of the imaginary particle moving in this potential, for several values of 
 $E$.  Again, shaded is the portion of the phase space where $q<0$ or $E>-1$.
Evolutions of the dimer are represented by the trajectories in 
the region that was left blank.

When $\gamma=0$, we have $r =const$ instead of Eq.\eqref{Z8}. In this case,
Eq.\eqref{Z13} is replaced with the pendulum equation:
\be
\frac{  {\dot \theta}^2 }{2\rho} + \sin \theta =E.
\label{Z20}
\ee
Here $E$ is determined by the initial conditions:
$E=\sin \theta(0)+ \frac12 Z^2(0)/ \rho(0)$.
The elliptic-function solution of \eqref{Z20} can be found in standard textbooks.
The pendulum is librating when $-1   \leq E   \leq 1$
and rotating when $E \geq 1$.
When averaged over a long interval, the $\theta$-coordinate of the rotating pendulum grows as a linear function of $t$.

\subsection{Fixed points} 
\label{Fixed_Points}

The fixed points of the two-dimensional system \eqref{Z10} satisfy
\be
2 \kappa^2 q=   \cos (q+  \chi).
\label{F1}
\ee
This equation has $2N+1$ roots, $q^{(0)}, q^{(1)}, ..., q^{(2N)}$, where $N$ depends on $\kappa$ and $\chi$.
For $2 \kappa^2>1$, there is only one point, of saddle type, irrespectively of the value of $0 \leq \chi < 2 \pi$. As $\kappa \to \infty$,
this fixed point is given by the asymptotic expression
\be
q^{(0)}=\frac{\cos \chi}{2 \kappa^2} 
- \frac{\sin 2\chi}{8 \kappa^4}+ O(\kappa^{-6}).
\ee

Assume now that the parameter $2 \kappa^2$ is being decreased from $1$.
For a generic value of
$\chi$,
a pair of new fixed points is born in a saddle-node bifurcation
 as $\kappa$ passes through $\kappa_n$ or ${\tilde \kappa_n}$,
  where $\kappa_n$ is defined as the root of the equation
\be
f(\kappa_n)= 2 \pi n-\chi, \quad n=1,2,...,
\label{F21}
\ee
 and ${\tilde \kappa_n}$ as the root of
\be
f({\tilde \kappa_n})=\chi+ \pi+ 2 \pi n, \quad
n=0,1,2,.... 
\label{F22}
\ee
Here
 the function
\[
f(\kappa)= \frac{1}{2 \kappa^2} \sqrt{1-4 \kappa^4} + \arcsin (2 \kappa^2)
\]
is monotonically decreasing from $\infty$ to $\pi/2$ as $2 \kappa^2$ grows from 0 to 1.
The equations \eqref{F21} and \eqref{F22} are arrived at 
by eliminating $q$ between \eqref{F1}
and the bifurcation condition 
\be
2 \kappa^2+ \sin(q+ \chi)=0.
\label{F20}
\ee

The values $\chi=\pi/2$ and $\chi=3\pi/2$ are 
 exceptional as these are associated with enhanced symmetry of Eq.\eqref{F1}.
 In each of these two cases both sides of Eq.\eqref{F1} are given by 
 odd functions of $q$; hence 
 {\it two} pairs of fixed points are born on crossing $\kappa_n$ or ${\tilde \kappa_n}$.
 (When $\chi=\pi/2$, we have $\kappa_{n+1}={\tilde \kappa_n}$; when 
 $\chi=3 \pi/2$, the correspondence is $\kappa_{n+2}= {\tilde \kappa_n}$.)

The order of the bifurcation points depends on $\chi$. We have
\[
 0< ...  \leq \kappa_3 \leq  {\tilde \kappa_2} \leq  \kappa_2   \leq {\tilde \kappa_1} \leq \kappa_1 \leq  {\tilde \kappa_0} \leq  \frac{1}{\sqrt{2}}
\]
when  $0 \leq \chi \leq \pi/2$;
\[
 0 < ...   \leq  {\tilde \kappa_2} \leq  \kappa_3   \leq {\tilde \kappa_1} \leq \kappa_2   \leq  {\tilde \kappa_0}  \leq  \kappa_1  \leq  \frac{1}{\sqrt{2}}
\]
when $\pi/2 \leq \chi \leq  3 \pi/2$, and, finally, 
\[
0< ... \leq  \kappa_4  \leq  {\tilde \kappa_1} \leq  \kappa_3   \leq {\tilde \kappa_0} \leq \kappa_2    \leq  \kappa_1  \leq  \frac{1}{\sqrt{2}}
\]
when 
$\chi \geq 3\pi/2$.

    It is important to emphasise that the phase portrait given in Fig.\ref{phase_portrait} cannot 
    be  simply ``wrapped" around the cylindrical surface in Fig.\ref{cylinder}.
    Different trajectories shown in  Fig.\ref{phase_portrait}  pertain to cylinders with different $\rho$;
    in particular, different fixed points belong to different cylinders.
    A natural question is how many fixed points lie on the surface of the cylinder of a given radius.

    To answer this,  we  
    calculate the value of the potential $V(q)$ at its points of  extrema:
    \[
    V(q^{(m)})= W(   \sin (q^{(m)}+\chi)            ), 
    \]
    where
  \be
 W(y)= -1+ \frac{(y+1)(y+ 4 \kappa^2-1)}{4\kappa^2}.
 \label{Q1}
\ee
When $2\kappa^2 \geq1$
(that is, when $\rho \leq  \gamma^2$), the function $W(y)$ lies above -1 when $y$ is 
in the interval $-1\leq y \leq 1$.
On the other hand, the maximum of the energy \eqref{J7},
attained at $\rho=1$, equals -1. 
Therefore, cylinders with $\rho \leq \gamma^2$ do not have fixed points
--- except  when $\rho=1$.
(This exceptional situation is obviously arising only if $\gamma>1$.)

 The cylinder with $\rho=1$ is special
 as it contains the origin $X=Y=Z=0$.
 When $\gamma>1$, the two-dimensional system \eqref{Z10}  with  
 $\rho=1$  and $\chi= 3\pi/2$
 has a saddle point at $q=0$.
  Unlike the saddles and centers in systems with other $\rho$ and $\chi$,
 this fixed point is accessible to the particle.
 In this case, Eq.\eqref{Z13} has the form 
 \be
 \frac{q_\tau^2}{2}  -  \frac{\gamma^2}{2}  q^2 + 2 \sin^2 \frac{q}{2}=0.
 \label{con}
 \ee
 The stable manifold of the saddle describes the solution of the
 dimer \eqref{coupler} with $|\psi_1|^2+|\psi_2|^2  \to 0$ as $z \to \infty$.
On the other hand,  the initial conditions $(q, q_\tau)$ constituting the unstable manifold 
 give rise to the blow-up regimes, $|\psi_1|^2+|\psi_2|^2  \to \infty$ as $z \to \infty$.

When $2 \kappa^2 \leq 1$ (i.e. $\rho \geq \gamma^2$), the function $W(y)$ has a (single) minimum,
at  $y=-2\kappa^2$, with 
\be
W_{\rm min} =W(-2\kappa^2)=-1-\frac{\gamma^2}{2 \rho} \left(1-\frac{\rho}{\gamma^2} \right)^2.
\label{Q2}
\ee
 Therefore, cylinders with  $W_{\rm min} < E $
will have two fixed points each. Substituting from \eqref{J7} for $E$, 
this inequality reduces to 
\be
(1-\gamma^2)(\rho-\gamma)>0.
\label{Q3}
\ee
When 
 $\gamma<1$, the inequality \eqref{Q3} requires $\rho>\gamma$.
 Here, the fixed points are at 
$\theta_{1,2}= \arcsin y_{1,2}$, where
\[
y_{1,2}= -\frac{\gamma^2} {\rho} 
\pm
\frac{\sqrt{(1-\gamma^2)(\rho^2-\gamma^2)}}{\rho}.
\]
The type of the fixed point --- considered as a fixed point 
of the imaginary particle --- is determined by
 the second derivative of $V(q)$:
\[
\frac{\partial^2 V}{\partial q^2}=-\frac{\gamma^2}{\rho} -\sin(q+\chi).
\]
Substituting $y_{1,2}$ for $\sin(q+\chi)$, we verify that $\theta_1$ is a saddle ($\frac{\partial^2 V}{\partial q^2}<0$) 
while $\theta_2$ is a centre ($\frac{\partial^2 V}{\partial q^2} >0$).

On the other hand,  when $\gamma>1$ the inequality \eqref{Q3} requires  $\rho < \gamma$.
However this is incompatible with  our assumption  $\rho>\gamma^2$, because
$\gamma^2$ becomes greater than $\gamma$ if $\gamma>1$.

  Since the system \eqref{Z10} is conservative, 
  each centre point is encircled by closed curves on the $(q, q_\tau)$ plane.
  Furthermore, it is not difficult to realise that each centre point is surrounded by closed oribts 
  on the cylindrical surface it belongs to (see Fig.\ref{cylinder}).
Indeed, let $V_\chi$ be the potential \eqref{V} corresponding to the parameter value $\chi$, 
denote $q^{(m)}(\chi)$  the corresponding  roots of Eq.\eqref{F1}  and
let $\rho(\chi)$ be the cylinder radius defined as a root of $V_\chi(q^{(m)}(\chi))=E$, with $E$ as in \eqref{J7}. 
Since 
\[
\left. \frac{\partial V_\chi(q)}{ \partial \chi} \right|_{q^{(m)}} = \cos (q^{(m)}+\chi) = 2 \kappa^2 q^{(m)}>0,
\]
 the value 
 $V_{\chi^\prime}(q^{(m)}(\chi))$, where $\chi^\prime=\chi+\delta \chi$ and  $\delta \chi<0$ is a small perturbation, 
 will be lower than $V_{\chi}(q^{(m)}(\chi))$ by a small amount.
 Therefore
  the conservation law \eqref{Z13} 
with $V_{\chi^\prime}$
will describe a periodic trajectory of  small radius on the  surface of the cylinder $\rho(\chi)$.
The trajectory will enclose the fixed point $q^{(m)}(\chi^\prime)$.

 In summary,  we need to distinguish between the situations with $\gamma<1$ and $\gamma>1$.
        When $\gamma<1$,
        cylinders of small radius $\rho< \gamma$ do not harbour any fixed points; all trajectories are unbounded.
        On the other hand, cylinders of  radius $\rho > \gamma$ feature two fixed points, a centre and a saddle;
        in this case periodic orbits 
  arise in addition to the unbounded motions. Finally,  there are no fixed points if $\gamma>1$. 
  (The only exception is the cylinder with $\rho=1$ which has  the saddle point at the origin, $r=0$.)
  All trajectories are spiralling up to infinity
  (except the stable manifold of the saddle at $\rho=1$).

After this paper has been submitted for publication, we have learnt of the preprint \cite{Pickton}
where the unboundedness of
trajectories for $\gamma>1$  was obtained within a different formalism.

 \subsection{Symmetry-broken phase}

The blowup of generic initial conditions in the symmetry-broken phase
($\gamma>1$) may be demonstrated without appealing to
details of the phase portrait. 
We now demonstrate this fact simply by 
considering the power imbalance between the 
two waveguides.

First, we show that in this symmetry-broken phase, 
 all initial conditions with $P_1>P_2$ blow up.
 From Eq.\eqref{coupler} it follows that 
\begin{align}
\frac{d}{dz} (P_1-P_2)
 = 2 (\gamma-1) (P_1+P_2)  \nonumber
\\
 +2(P_1+P_2) + 2i (\psi_1^*\psi_2-\psi_1 \psi_2^*)
  \nonumber                       \\
 \geq 2(\gamma-1) (P_1 +P_2) + 2(\sqrt{P_1}-\sqrt{P_2})^2,
  \label{b1}   
\end{align}
whence
\be
\frac{d}{dz} (P_1-P_2) \geq 2(\gamma-1) (P_1-P_2). 
\nonumber
\ee

By the Gronwall inequality, the difference $P_1-P_2$ 
tends to infinity for any
initial conditions with 
$P_1(0) >P_2(0)$. That is,  any initial conditions with 
$P_1(0) >P_2(0)$ lead to a blow-up.

 Most of solutions with $P_2(0)  \geq P_1(0)$
will also blow up.
To show this, we first 
observe that
Eq.\eqref{b1} implies
\begin{align}
\frac{d }{d z}  (P_2-P_1)  \nonumber \\   \leq  
 -2(\gamma-1)(P_1+P_2) 
 -2(\sqrt{P_2}-\sqrt{P_1})^2
<0.   \label{q1}
\end{align}
According to \eqref{q1}, 
the quantity $P_2 - P_1$ must
decrease  until $P_2 = P_1$. 
If $P_1$ and $P_1$ are not zero at the moment when they become
equal, the difference $P_2-P_1$ will continue to decrease.
Once the difference $P_2-P_1$ has become negative, 
the system is in the blowup regime described above.

The quantities $P_1$ and $P_2$ may simultaneously go to zero
only if $\rho = 1$. 
Indeed, the product $P_1P_2$ equals $X^2+Y^2$
while 
\[
X^2+Y^2=(\rho-1)^2+ 2 \rho(1+ \sin \theta) \geq (\rho-1)^2;
\]
hence 
$ P_1 P_2 \geq (1- \rho)^2$.
The trajectory with $P_1$, $P_2 \to 0$ 
is the stable manifold of the saddle point $\rho=1$, $\theta=3\pi/2$, $Z=0$
(that is, of the point $X=Y=Z=0$).

In conclusion, in the symmetry-broken phase ($\gamma>1$),
all initial conditions lead to the blowup of solutions,
except initial conditions that lie on the stable manifold of the 
saddle point $\psi_1=\psi_2=0$.

\section{$\mathcal{AC}$ dimer} 
\label{sAC}


The $\mathcal{AC}$  coupler offers an alternative to the $\mathcal{PT}$-symmetric configuration
of gain and loss   \cite{ABRF}.
 The
arrangement consists of two lossy waveguides placed in
an active medium. Instead of providing power gain in the
core of (one of the) waveguides, the structure boosts the evanescent
fields which couple the two channels due to their close
proximity.

The optical field in the two guides is described by the 
amplitudes $\psi_1$ and $\psi_2$. These satisfy
\begin{subequations}
\label{B1}
\begin{align}
i \frac{d \psi_1}{dz}+  \beta |\psi_1|^2\psi_1  +\psi_2 
=- i \gamma  \psi_1 
+   ia \psi_2,  
\label{B1-2}
\\
 i\frac{d \psi_2}{dz} + \beta |\psi_2|^2\psi_2 + \psi_1 
 =-i \gamma  \psi_2
 + ia \psi_1.
\label{B1-1}
\end{align}
\end{subequations}
Here $a$ and $\gamma>0$ are the gain and loss coefficient, respectively.
We assume $a>\gamma$ (because if $a<\gamma$, all solutions decay to zero
\cite{ABRF}).
The coefficient $\beta$ measures the strength of nonlinearity.
The choice $\beta>0$ corresponds to the focusing and $\beta<0$ to defocusing nonlinearity.

We note that a closely related system, with $\beta<0$,
describes radiative coupling and weak lasing of exciton-polariton condensates
\cite{Aleiner}. 
Unlike the $\mathcal{PT}$-symmetric dimer, the $\mathcal{AC}$ couplers with the opposite sign
of $\beta$ are not equivalent. The staggering transformation \eqref{stag} 
changes   the sign of $a$ in addition to 
the sign of the nonlinear term. If the sign of $a$ is fixed by the condition $a> \gamma>0$,
the cases $\beta>0$ and $\beta<0$ have to be considered independently. 


Linearising \eqref{B1} about $\psi_{1,2}=0$ one checks that the symmetric 
part of the small perturbation,  $u=\psi_1+\psi_2$, gains energy and grows:
\[
u(z)= u(0) e^{(i+a-\gamma)z}.
\] On the other hand, 
the antisymmetric normal mode, $v=\psi_1-\psi_2$, loses energy and
decays to zero:
\[
v(z) =v(0)e^{-(i+a+\gamma)z}.
\]
The numerical evidence \cite{ABRF} is that the nonlinearity
which couples the two modes, may drain the energy gained by the 
symmetric mode through the antisymmetric channel.
Below, we study the blowup arrest analytically, and 
identify the type of nonlinearity capable of this job.

The system \eqref{B1} has two invariant manifolds. One is defined by the 
reduction $\psi_1=\psi_2 \equiv \psi$, where $\psi$ satisfies
\be
i \frac{d \psi}{dz}+  \beta |\psi|^2\psi +\psi
= i (a-\gamma)   \psi,
\label{V1}
\ee
and the other one by 
$\psi_1=-\psi_2 = \psi$, where 
\be
i \frac{d \psi}{dz}+  \beta |\psi|^2\psi - \psi
= -i (a+ \gamma)   \psi.
\label{V2}
\ee
All solutions of \eqref{V2} decay to zero; letting 
$|\psi(0)|^2=A^2$, we have
\be
\psi(z)=\psi(0)
e^{  -(a+\gamma)z +i(\beta A^2-1)z}.
\label{W10}
\ee
On the other hand, all solutions of \eqref{V1} blow up, exponentially: 
\be
\psi(z)=\psi(0) e^{  (a-\gamma)z + i(\beta A^2+1) z}.
\label{V3}
\ee
The issue we are exploring in what follows, is whether 
initial conditions that lie close to the ``blow-up manifold"
$\psi_1=\psi_2$ blow up as well.

Performing the polar decomposition of the fields $\psi_1= \sqrt{P_1} e^{i \Phi_1}$
and $\psi_2= \sqrt{P_2} e^{i (\Phi_2)}$, one checks that 
$\Phi_1$  can be separated from the other three variables. 
This phase variable satisfies
\[
{\dot \Phi_1}= \frac{\beta (r+X)}{2} + \frac{Z+ aX}{2(r+X)},
\]
while the remaining equations of motion can be written as
\begin{subequations}
\label{E1} 
\begin{align}
{\dot X}=-\gamma X  - Y,   \label{C1}  \\
{\dot Y}=-\gamma Y +  X -  \beta XZ,    \label{C2}   \\
{\dot Z}=-\gamma Z + a r + \beta XY.    \label{C3}  
\end{align} 
\end{subequations} 
Here $X=\frac12(|\psi_1|^2-|\psi_2|^2)$ measures the power imbalance between the two waveguides;
 $Y=\frac{i}{2} (\psi_1\psi_2^*-\psi_1^*\psi_2)$
characterizes the energy flux from the first to the second channel, and
$2aZ$ --- where  $Z=
\frac12( \psi_1\psi_2^*+\psi_2\psi_1^*)$ ---
is the total gain in the system. The Stokes variables $X,Y$, and $Z$  are three components of the vector ${\bf r}$,  with  $r=\sqrt{{\bf r}^2}=\frac 12 (P_1+P_2)$.
[Note that the Stokes variables have been introduced differently from \eqref{change}; this is done in order 
to elucidate parallels in the geometry of the phase spaces of the two systems.]
The overdot indicates differentiation with respect to the fictitious time variable,
$t=2z$, which we introduce for convenience of analysis.

\subsection{Defocusing nonlinearity}

With the $\mathcal{AC}$ dimer being only recently introduced, its 
phenomenology still needs to be elucidated.
One issue that requires a careful investigation is the type of nonlinearity 
that is necessary for the operation of the structure as an optical coupler.
The choice of the self-focussing Kerr nonlinearity in the original version of this
structure \cite{ABRF} was arbitrary; 
the defocussing nonlinearity could have been an equally acceptable candidate.

In this subsection we show, however,  that  the defocussing cubic nonlinearity ($\beta<0$)
is unable to prevent the blow-up. 

Our analysis makes use of the function
\be
{\mathcal L}= Z+ \frac{\beta}{2}X^2,
\label{L}
\ee
which 
satisfies
\be
{\dot {\mathcal L}} =
  ar-\gamma Z- \gamma \beta X^2.
\label{Lt}
\ee

We start by considering the initial conditions
$X, Y, Z$
such that   ${\mathcal L} \leq 0$.
From the definition of $\mathcal{L}$ we have 
$ -\beta X^2 \geq 2Z$. Using this inequality in \eqref{Lt} we obtain
\[
{\dot {\mathcal L}} \geq ar+\gamma Z \geq (a-\gamma)|Z|.
\]
This means that ${\mathcal L}(t)$ will either grow until it is positive,
or tend to zero as $t   \to \infty$. The latter is only possible if the initial
condition lies on the 
stable manifold of the origin [described by Eq.\eqref{W10}].

Thus we need to consider only initial conditions satisfying ${\mathcal L}(0)>0$.
When $\beta<0$, Eq.\eqref{Lt}  implies
\be
{\dot {\mathcal L}}   \geq   (a-\gamma)r.
\label{ineq}
\ee
Since $\mathcal{L} \leq r$
and so ${\dot {\mathcal L}}  \geq (a-\gamma) {\mathcal L}$,
the Gronwall inequality gives
${\mathcal L}(t) \geq {\mathcal L}(0) e^{(a-\gamma) t}$
for any initial conditions with ${\mathcal L}(0)>0$.
This means that these initial conditions 
blow up: $|\psi_1|^2+|\psi_2|^2 \to \infty$ as $z \to \infty$.
(From the structure of $\mathcal{L}$ it follows that 
  $|\psi_1|$ and $|\psi_2|$ grow to infinity
at the same rate.)

Thus the defocusing nonlinearity cannot arrest the blowup of 
solutions of the linear ${\mathcal AC}$-dimer.
In what follows we concentrate on the 
 focusing  case ($\beta > 0$) 
and scale $\psi_{1,2}$ so that $\beta=1$. 

\subsection{Instability of the blowup solution}

Here our purpose is to explore trajectories that start in the vicinity of the 
blow-up manifold \eqref{V3}.
In terms of $X,Y$ and $Z$, this manifold
is given by the positive vertical axis: $X=Y=0$; $Z>0$. We 
wish to determine whether these trajectories escape to infinity 
or remain in the finite part of the space.

In terms of the cylindrical coordinates Eqs.\eqref{E1} acquire the form 
\begin{subequations}
\label{E2} 
\begin{align}
{\dot \rho}=  \left[ -\gamma  - \frac12 Z \sin(2 \theta)  \right] \rho,   \label{D1}  \\
{\dot Z}=-\gamma Z + a \sqrt{\rho^2+Z^2} +\frac12 \rho^2 \sin (2 \theta),  \label{D2}  \\
{\dot \theta}=1- Z \cos^2 \theta.   \label{D3}  
\end{align} 
\end{subequations} 
Here $X=\rho \cos \theta$ and $Y=\rho \sin \theta$.

We assume that the motion starts in a narrow cylinder around the 
$Z$ axis, and linearise in small $\rho$.
Equation \eqref{D2} is then simply 
${\dot Z}=(a-\gamma) Z$, so that $Z$ grows:
$Z(t)=Z(0)e^{2 \lambda t}$, where $2\lambda=a-\gamma$.
Assume that  $Z(0)>1$ while $\theta(0)$  is in the vicinity of $\pi/2$ or $-\pi/2$.
Writing $\theta=\pm \pi/2+ \epsilon(t)$, Eq.\eqref{D3} becomes
\[
{\dot \epsilon}=1- Z(t) \epsilon^2,
\quad
Z=Z(0)e^{2 \lambda t}.
\]
The solution of this Riccati equation is
\be
\epsilon=\frac{\lambda}{Z} 
+ \lambda
\frac{d}{dZ} \ln \left[ K_1 \left(  \frac{\sqrt{Z}}{\lambda}\right)+ C I_1 \left(  \frac{\sqrt{Z}}{\lambda}\right) \right]^2,
\label{G1}
\ee
where $I_1(w)$ and $K_1(w)$ are the modified Bessel functions of order one, and 
$C$ is a constant of integration.
As $Z$ grows, \eqref{G1} gives $\epsilon \to \pm 1 / \sqrt{Z}$,
where the top respectively bottom sign results from 
choosing $C \neq 0$ respectively $C=0$. 

When $\theta$ approaches $\pi/2$ from below ($\epsilon<0$)
or $-\pi/2$ from above ($\epsilon>0$),
the contents of the square bracket in \eqref{D1} tends
to $-\gamma -\sqrt{Z}$. The radius 
$\rho(t)$ continues to decrease
while $Z$ continues to grow. The trajectory 
is captured in a blowup regime.

On the other hand, when $\theta$ tends to $\pi/2$ from above
or $-\pi/2$ from below,
 the square bracket becomes $\sqrt{Z}-\gamma$,
which is large and positive.
The radius $\rho$ then  starts increasing as a double exponential function,
and
the last, negative,  term in \eqref{D2} outgrows the first two terms.
This suppresses any further growth of $Z$;
the trajectory moves away from the blowup manifold [Fig.\ref{cylinder}(b)].

When $\rho(0)$ is small, a tiny perturbation is sufficient to  change the sign of $\epsilon(0)$
and divert the phase point from a trajectory escaping to infinity. 
Therefore,  even though there are 
trajectories with $X,Y \to 0$, $Z \to \infty$ as $t \to \infty$,
these blowup solutions are unstable
and will not be observed in any practical situation. 
(In particular, the blowup cannot be observed in numerical simulations 
of the $\mathcal{AC}$ dimer.)

\section{Conclusions}
\label{Conclusions}

In the case of the $\mathcal{PT}$ symmetric dimer, our results include the following.


(1)
In the symmetry-broken phase ($\gamma>1$) we have demonstrated that all
initial conditions   (except initial conditions from a special degenerate class) blow up.

(2)
We have elucidated the geometry of the phase space of the dimer. 
In particular, we have shown that the phase space is foliated into coaxial two-dimensional cylinders.
Cylinders of small radius only harbor trajectories that escape to infinity; 
these describe the blow-up regimes of the dimer. When $\gamma \leq 1$, 
cylinders with larger radii host periodic trajectories in addition to the unbounded motions.

An  implication of  the  phase space foliation 
and the conservativity of motion  on each cylinder, is that 
the  blow-up regime is stable.  Small perturbations 
 may shift  the phase point around the cylindric surface, or push it
  from one cylinder to another, but this will not take it to the  bounded trajectories.
  The evolution carries the phase point further away from the domains
  of finite motion.

For the $\mathcal{AC}$ dimer, we have established that

(1)
The defocusing Kerr nonlinearity is unable to suppress the blowup. Generic 
initial conditions lead to unbounded trajectories.

(2)
The phase space of the $\mathcal{AC}$ dimer
is genuinely three-dimensional 
and not foliated. When the nonlinearity is focussing, there is a
domain of initial conditions occupying nonzero phase volume,
 that lead to
blow-up regimes. However all the unbounded trajectories
lie within a rapidly narrowing  funnel centered on the vertical axis.
The blow-up funnel is unstable:  a small perturbation is sufficient
to kick a trajectory out of the funnel
and send it towards a stable limit cycle.

In conclusion, the same, focussing cubic, nonlinearity plays a dramatically 
different role in the dynamics of the $\mathcal{PT}$ and $\mathcal{AC}$ dimer.
In the $\mathcal{PT}$-symmetric arrangement of the gain and loss, 
the nonlinearity promotes the blowup of solutions. 
In the case of the $\mathcal{AC}$ coupler,  the nonlinearity suppresses the 
blowup by coupling the linearly excitable to the linearly damped normal mode.
We have shown that this opposite effect of the nonlinearity is
due to the difference in the geometry of the phase space of the two systems.

\section*{Acknowledgments}

The project was supported by the NRF of South
Africa (Grants UID 85751 and 78950).
We acknowledge instructive conversations with  N.  Akhmediev, M.  Gianfreda,  V.  Konotop,
D.  Skryabin,  A. Smirnov, and M. Znojil.
We are grateful to the referee for bringing Ref.\cite{Pickton}  to our attention.


\section*{References}
\begin{thebibliography}{99}











\bibitem{PT_breaking}
S. Klaiman, U. G\"unther, and N. Moiseyev, Phys. Rev. Lett.  {\bf 101}
080402 (2008);
Z. Lin, 
H.  Ramezani, T.  Eichelkraut,   T.  Kottos,  H.  Cao, and D.  N. Christodoulides,
Phys. Rev. Lett.  {\bf 106}  213901 (2011);  
A. Regensburger,
C.  Bersch, M.-A. Miri, G. Onishchukov, D. N. Christodoulides, and
U. Peschel, Nature {\bf 488} 167 (2012)



\bibitem{SXK} A.A. Sukhorukov,  Z.Y. Xu, Yu.S. Kivshar, 
Phys. Rev. A {\bf 82} 043818 (2010)



\bibitem{Rueter}
C. E. R\"uter, K. G. Makris, R. El-Ganainy, D.N. Christodoulides, M. Segev, and D. Kip, 
Nat. Phys. {\bf 6} 192 (2010);  T. Kottos,  {\it ibid.}  166.


\bibitem{Musslimani}
K. G. Makris,  R. El-Ganainy, D. N. Christodoulides,  and Z. H. Musslimani, 
Phys. Rev. Lett. {\bf 100} 103904 (2008)

\bibitem{Zheng} M. C. Zheng,
 D. N. Christodoulides,  R. Fleischmann, T. Kottos, 
Phys. Rev. A {\bf 82} 010103 (2010)


 \bibitem{Longhi} S. Longhi, Phys. Rev. Lett. {\bf 103}  123601 (2009)

\bibitem{Guo} A. Guo, G. J. Salamo, D. Duchesne, R. Morandotti, M. Volatier-Ravat, 
V. Aimez, G. A. Siviloglou, and D. N. Christodoulides.
  Phys. Rev. Lett. {\bf 103} 093902 (2009)

\bibitem{Ramezani} H. Ramezani,  T. Kottos, V. Kovanis, and D. N. Christodoulides, 
Phys. Rev. A {\bf 85} 013818 (2012)


\bibitem{BSSDK} I. V.Barashenkov, S. V. Suchkov, A. A. Sukhorukov, S. V. Dmitriev, and Y. S. Kivshar,
 Phys. Rev. A {\bf 86}  053809 (2012)



\bibitem{RKEC} H. Ramezani,
 T. Kottos,  R. El-Ganainy, and D.H. Christodoulides, 
Phys. Rev. A {\bf 82} 043803 (2010)







\bibitem{Recent_PT}  
  S. Hu and W. Hu,
 J. Phys. B: At. Mol. Opt. Phys. {\bf 45}   225401  (2012); 
 Y.  He and D.  Mihalache, Phys. Rev. A  {\bf 87}    013812   (2013); 
 Y. V.  Bludov,  V.  V. Konotop,  B.  A. Malomed, 
Phys. Rev. A {\bf 87}    013816   (2013);     
G. Della Valle, S. Longhi, Phys. Rev. A {\bf 87} 022119 (2013);
K. Li, D. A. Zezyulin, V. V. Konotop, P. G. Kevrekidis, Phys. Rev. A {\bf 87} 033812 (2013);
X. L. Shi, F. W. Ye, B. Malomed, X. F. Chen, Opt. Lett. {\bf 38} 1064 (2013);
M. Duanmu, K. Li, R. L. Horne, P. G. Kevrekidis, N. Whitaker,
Phil. Trans. Roy. Soc. A - Math. Phys. Eng. Sci. {\bf 371} 20120171 (2013);
Y. V. Bludov, R. Driben, V. V. Konotop, B. A. Malomed, Journ. Optics {\bf 15} 064010 (2013);
S. Nixon, J. K. Yang, Optics Lett. {\bf 38} 1933 (2013);
B. Peng,  \d{S}. K. \"{O}zdemir, F. Lei, F. Monifi, M. Gianfreda, G. L. Long, S. Fan, F. Nori, C. M. Bender, L. Yang,
submitted to Nature (2013)



\bibitem{ABRF} N. V. Alexeeva, I. V. Barashenkov, K. Rayanov, and S. Flach, arXiv: 1308.5862






\bibitem{CSP}
Y. Chen, A. W. Snyder, and D. N. Payne,
IEEE Journ.  Quant.  Electronics {\bf 28}   239 (1992)



\bibitem{finite}
K.  Li and P.  G. Kevrekidis, Phys. Rev. E {\bf 83}  066608 (2011);
J.  D'Ambroise, P.  G. Kevrekidis, and S.  Lepri,  J. Phys. A Math. Theor. {\bf 45} (2012) 444012;
D.  A. Zezyulin and V. V. Konotop, Phys. Rev. Lett. {\bf 108}  213906
(2012); K. Li, P. G. Kevrekidis, B. A. Malomed, and U. G\"unther,
J. Phys. A:  Math. Theor. 45, 444021 (2012);
 I. V. Barashenkov, L. Baker, N. V. Alexeeva,
Phys. Rev. A {\bf 87}   033819 (2013);
P. G. Kevrekidis, D. E. Pelinovsky, and D. Y. Tyugin, J. Appl. Dynam. Syst. {\bf 12} 1210 (2013)




\bibitem{infinite}
S. V. Dmitriev, A. A. Sukhorukov,  and Y. S. Kivshar, 
Opt. Lett. {\bf 35}  2976 (2010);
S. V. Suchkov, B.  A. Malomed, S. V. Dmitriev, and Y. S.Kivshar,
Phys. Rev. E {\bf 84} 046609 (2011); R. Driben and B. A. Malomed, Opt. Lett. {\bf 36}  4323 (2011);
 S. V. Suchkov, A. A. Sukhorukov, S. V. Dmitriev, Y. S. Kivshar, EPL {\bf 100}  54003 (2012);
D. E. Pelinovsky, P. G. Kevrekidis, D. J. Frantzeskakis, EPL {\bf 101} 11002 (2013)


\bibitem{KPT}
P. G. Kevrekidis, D. E. Pelinovsky, and D. Y. Tyugin,
J. Phys. A: Math. Theor. {\bf 46} 365201 (2013)








\bibitem{Aleiner}
I.  L. Aleiner,  B.  L. Altshuler, and Y.  G. Rubo. Phys. Rev.  B {\bf 85} 121301 (2012)


\bibitem{Pickton}
J. Pickton and H. Susanto, arXiv: 1307.2788





\end{thebibliography}
\end{document}